\documentclass[%
 reprint,
 amsmath,amssymb,
 aps,
prl,citeautoscript,twocolumn
]{revtex4-2}

\usepackage{braket}
\usepackage{caption}

\usepackage{graphicx}
\usepackage{dcolumn}
\usepackage{bm}
\usepackage{hyperref}

\begin{document}


\title{Structural Dynamics and Strong Correlations in Dynamical  Quantum Optical Lattices}

\author{Adrián U. Ramírez-Barajas}
\affiliation{%
Instituto de Física, LSCSC-LANMAC, Universidad Nacional Autónoma de México, Ciudad de México 04510, Mexico
}%

\author{Santiago F. Caballero-Benitez}%
 \email{scaballero@fisica.unam.mx}
\affiliation{%
Instituto de Física, LSCSC-LANMAC, Universidad Nacional Autónoma de México, Ciudad de México 04510, Mexico
}%

\makeatletter
\renewcommand\@biblabel[1]{\textsuperscript{#1.}}
\makeatother

\begin{abstract}
When placing an ultracold atomic gas inside a cavity, the light-matter coupling is enhanced and nonlinear atomic
dynamics are generated, offering a promising platform for quantum simulation of models with short- and  
long-range interactions. Recently, superradiant self organized phases for ultracold atomic gases inside a cavity, pumped by a blue detuned optical lattice,
have been observed. Here, we explore the formation of quantum many-body phases with strongly interacting bosonic atoms inside an optical cavity,
subject to transverse blue detuned pumping. We analyze the interplay between superradiant self-organization with  superfluid and Mott insulator phases, without the need 
of including higher lying bands, as the Wannier functions are dynamically linked to the cavity light via backaction. We observe different kinds of structural phase transitions driven by the light inside the cavity and the interplay with atomic collisions. We observe the mode softening at the critical points in the quantum phase transitions which can be measured in future experiments.
\end{abstract}


\date{\today}                              
\maketitle


Quantum many body phases  in the strongly correlated  regimen can be studied by means of ultracold gases inside an optical lattice, where  the lattice structure is dictated by the external trapping potential~\cite{book:92350169}. 
By placing the atomic gas inside a high finesse optical cavity, strong coupling between matter and the quantized cavity field modes is achieved.
As the atoms  scatter light collectively to the cavity,  the lattice potential is modified and self consistent light matter states can be formed, allowing   the emergence of dynamical lattice potentials~\cite{RevModPhys.85.553,doi:10.1080/00018732.2021.1969727,PhysRevA.93.063632,PhysRevLett.115.243604,PhysRevLett.128.080601}.  This permits the study of structural phase transitions between different lattice configurations~\cite{PhysRevResearch.3.L012024, Shakya_2023}, which have been studied
using several pumps\cite{Shakya_2023}, multiple cavity modes \cite{Gopalakrishnan2009} or coupling to the two quadratures of a single cavity mode~\cite{PhysRevResearch.3.L012024, Dreon2022}. Light-mediated interactions are useful for quantum simulation of models with short- and long-range effects, as all atoms are coupled to all the other within the cavity \cite{RevModPhys.85.553,doi:10.1080/00018732.2021.1969727}. Quantum uncertainties of the potential and the possibility of atom-field entanglement~\cite{PhysRevA.89.023832} lead to modified phase transition characteristics, the appearance of new phases or even quantum superpositions of different phases \cite{Maschler2008,PhysRevLett.115.243604}, e.g. the superradiant (SR) Mott insulator (MI)~\cite{Hemmerich} or the supersolid phase~\cite{Landig2016}. This is the case of the self-organization of a BEC inside a cavity and pumped by a  blue detuned optical lattice, in which the atoms are dragged to the intensity minima of the external light field.
Despite the energetic cost  of an additional repulsive potential caused by atomic self-organization, destructive interference occurs at the
position of the atoms,  carving out parts of the repulsive
pump lattice potential and lowering the potential energy \cite{PhysRevLett.123.233601}. In this scenario, dimensional crossover can occur, enhancing the variety of the quantum systems that can be simulated by ultracold atoms \cite{Shakya_2023} and allowing the formation of striped phases \cite{PhysRevA.94.033607,Shakya_2023}. 

In typical setups exploring structural changes in ordinary matter~\cite{Structural1,StructuralB}, classical phase transitions can present mode softening in second order phase transitions, while there is an absence or weak mode softening in first order transitions as a function of temperature. Mechanisms for these behaviour are tightly linked with the anharmonic change of atomic displacements and the modification to the spatial symmetry. In a quantum system at low temperatures the situation is complicated by the fact that now besides the structural change, quantum phase transitions~\cite{Sachdev_2011} take place at $T=0$ while quantum degeneracies and symmetry breaking also determine the macroscopic state of the system. Implicit anharmonicities emerge due to quantum many-body interactions. An excellent platform to explore these mechanisms is posed by ultracold systems with optical lattices  inside  high-Q cavities, where the interplay between the lattice structure, many-body interactions and emergent order can lead to structural phase transitions and mode softening~\cite{Landig2015}.

\begin{figure*}
    \centering
    \includegraphics{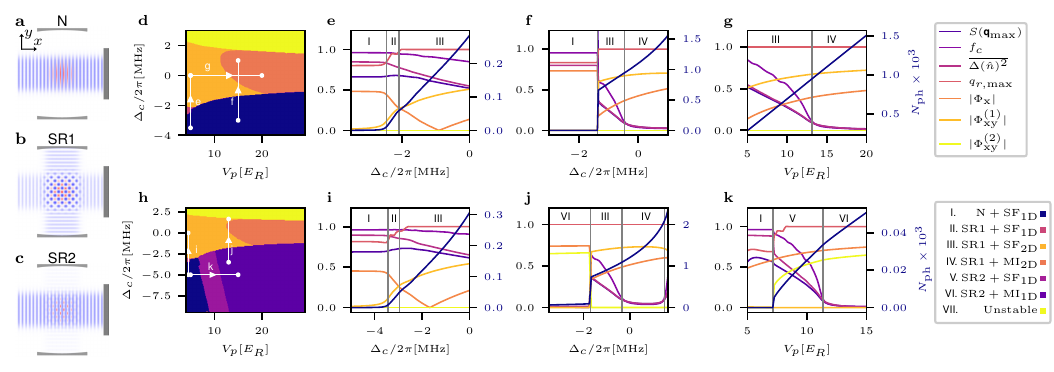}
    \caption{
     \textbf{Quantum phases of interacting ultracold atoms in a blue detuned quantum optical lattice.}
     \textbf{a}-\textbf{c}, Experimental scheme.  The atomic cloud (red) is placed inside an optical cavity (curved mirrors on the top and bottom) aligned  along the $y$ axis. The cloud is pumped along the $x$ axis by a retroreflected blue detuned laser field, which induces a repulsive optical lattice (blue), repelling the atoms from the maxima of the light field. 
    \textbf{a}, Normal phase N, with no scattering into the cavity mode.
    \textbf{b}, \textbf{c}, Atomic self organization and collective scattering of photons into the resonator mode can become energetically favorable.
    Two superradiant phases (\textbf{b}) SR1 and (\textbf{c}) SR2  can emerge in our system, each one coupled to a different quadrature of the cavity field and with its own lattice structure.  
     \textbf{d}, \textbf{h}, Phase diagrams for the balanced (\textbf{d}) ($g_\mathrm{2D}/(E_Rk_p^{-2})=1$, $\gamma=1$) and imbalanced (\textbf{h}) ($g_\mathrm{2D}/(E_Rk_p^{-2})=0.5$, $\gamma=1.37$) pump cases. 
        Different trajectories (white) in the phase diagrams correspond to (\textbf{e}-\textbf{g}) and (\textbf{i}-\textbf{k}). Panels \textbf{e}-\textbf{g}, \textbf{i}-\textbf{k}, we show the maximum of the static structure factor $S(\textbf{q}_\mathrm{max})$,  its radial  position  $q_{r,\mathrm{max}}$, the condensate fraction $f_c$, the mean of the on-site density fluctuations $\overline{ \Delta(\hat{n})^2} $, the cavity mode photons $N_\mathrm{ph}$ and the structural order parameters $|\Phi_\mathrm{x}|$, $|\Phi_{\mathrm{xy}}^{(1)}|$ and $|\Phi_{\mathrm{xy}}^{(2)}|$. The dominant structural order parameter determines the structural phase 1D or 2D depending on the quadrature.
      The phase boundaries are given by the gray vertical solid lines. 
      (Right lower panel) Dictionary of  the phases in the system.  Increasing $\Delta_c$, an unstable phase appear in both cases (light yellow zone in \textbf{d} and \textbf{h}), where there is no stationary solution for the light field amplitude inside the cavity  $|\alpha|$. The results were obtained using our light-matter DRMG method.
    }
    \label{fig:Fig1}
\end{figure*}

 In this work, we study the self-organization of strong interacting bosonic atoms inside a cavity, transversely illuminated by a blue detuned pump, similar to \cite{PhysRevLett.123.233601,PhysRevResearch.3.L012024}. In similar configurations, at the level of mean-field, atomic interaction have been linked  the  formation of higher orbital phases\cite{PhysRevA.106.023315} of the underlying lattice potential.
Here,  we map the lattice problem to a dynamical lattice model, where the Wannier functions are dynamically dependent on the full pump-cavity field, allowing us to describe the superradiant superfluid and Mott insulating phases, without the need to include additional excited bands.
Our approach allows us to simulate the system in a self-consistent dynamical framework while considering the effects of strong quantum correlations. In our theory, the light field is represented as a coherent field consistent with experiment and the atomic dynamics in 2D are simulated using Density Matrix Renormalization Group (DMRG)~\cite{DMRG1,RevModPhys.77.259,DMRG3,itensor1,itensor2} in a full self-consistent approach where the Wannier functions are dependent on the light field amplitude parametrically, see End Matter. Similar approaches combining mean field theory with DMRG have being explored recently  to describe dimensional crossovers with strong quantum correlations~\cite{PhysRevB.102.195145,PhysRevX.13.011039,10.21468/SciPostPhys.15.6.236}, while other Monte-Carlo based and hybrid methods have been successfully employed~\cite{10.21468/SciPostPhys.15.2.050,PhysRevResearch.7.013021} and related to experiments~\cite{DimXExp1,DimXExp2,PhysRevResearch.5.013136}.

We consider bosonic atoms at $T=0$ illuminated in the $x$ direction by a retro-reflected stationary pump field with frequency $\omega_p$ and wave vector $\textbf{k}_p=(2\pi/\lambda_p)\textbf{e}_x$,  coupled to the mode of a high-Q optical cavity with frequency $\omega_c$ and wave vector $\textbf{k}_c=(2\pi/\lambda_p)\textbf{e}_y$. We work in the natural units of the recoil energy $E_R$.
The pump is blue detuned with respect to the atomic resonance $\omega_a$,  $\Delta_a=\omega_p-\omega_a>0$, which causes the atoms to be repelled from the intensity maxima of the light field. 
Depending on the focus point of the incident pump beam, a balance parameter $\gamma=\sqrt{E_+/E_-}$ can be introduced, were $E_+$ and $E_-$ are the   incident and retro-reflected pump beams amplitude. This leads the system to couple to the two orthogonal quadratures ( $ Q$ and $ P
$) of the cavity field \cite{PhysRevResearch.3.L012024}, see End Matter.  The system scheme is shown in Fig.  \ref{fig:Fig1} a-c.
The enhanced cavity coupling of the light with atoms can induce self-organized fields that indicate the superradiant order the system can support,  $\hat\Theta_1=\int d^2 r\cos(\textbf{k}_p\cdot\textbf{r})\cos(\textbf{k}_c\cdot\textbf{r})\hat n(\textbf{r})$ related to the $Q$ and $\hat\Theta_2=\int d^2 r  \sin(\textbf{k}_p\cdot\textbf{r})\cos(\textbf{k}_c\cdot\textbf{r})\hat n(\textbf{r})$ related to the $P$ light field quadratures in the cavity with the density operator $\hat n(\textbf{r})= \hat \Psi^\dagger(\textbf{r}) \hat \Psi(\textbf{r})$ and  the atomic field operator $\hat \Psi(\textbf{r})$ that represents the atoms in the system.

 Structural phases can be characterized by the structural order parameters
     $|\Phi_\mathrm{xy}^{(1)}| = |\braket{\cos(k_p x) \cos(k_p y)}|$,
    $ |\Phi_\mathrm{xy}^{(2)}| = |\braket{\sin(k_p x) \cos(k_p y)}|$ and
      $ |\Phi_\mathrm{x}| = |\braket{\cos(2 k_p x) }|$,
 which quantify the localization of the wave function around the antinodes of the respective spatial terms, which have different symmetries. 
 The dominant order parameter determines the structural phase.
 $|\Phi_\mathrm{xy}^{(1)}|$ describes a two dimensional lattice with $\lambda_p$-periodic spatial order \cite{Nagy2008}, where the atoms localize around the antinodes of the emergent interference term $\cos(k_p x) \cos(k_p y)$ originated by  the coupling to the $Q$ quadrature of the cavity field.  Similarly, $|\Phi_\mathrm{xy}^{(2)}|$ is related to the 
 interference term $\sin(k_p x) \cos(k_p y)$ corresponding to the emergent $P$ quadrature cavity field.  When the contribution
from the pump field to the optical lattice dominates
over the other emergent terms, with weak periodic modulation
along $y$ and strong periodic modulations along $x$, the structural phase is
characterized by $|\Phi_\mathrm{x}|$.

In Fig. \ref{fig:Fig1} d, h, we show the ground state phase diagrams for  the balanced ($\gamma=1,\ g_\mathrm{2D}/(E_Rk_p^{-2})=1$) and the imbalanced   ($\gamma=1.37,\ g_\mathrm{2D}/(E_Rk_p^{-2})=0.5$) systems, the effective interaction strength due to atomic collisions is $g_\mathrm{2D}$.
Similarly to previous studies\cite{PhysRevResearch.3.L012024}, we identify two distinct superradiant phases SR1 and SR2, with $\braket{\hat\Theta_1}\neq 0$ and $\braket{\hat\Theta_2}=0$, or $\braket{\hat\Theta_1}= 0$ and $\braket{\hat\Theta_2}\neq0$ respectively, for which the cavity light amplitude $\alpha=\langle\hat a\rangle\neq 0$. The two superradiant phases can be distinguished by the corresponding  cavity light field phase $\mathrm{arg}(\alpha)$, which is $n\pi$ for the SR1 and $(n+1/2)\pi$ for the SR2, with $n$ an integer. 
We found the following phases: (I) a superfluid striped phase with no light scattering (normal phase N) N+SF$_\mathrm{1D}$; (II)   a superfluid striped phase  with  light scattering ($Q$ quadrature)  SR1+SF$_\mathrm{1D}$;  (III) a superfluid phase with light scattering ($Q$ quadrature) in a 2D square lattice SR1+SF$_\mathrm{2D}$; (IV) an insulator phase with light scattering ($Q$ quadrature) in a 2D square lattice SR1+MI$_\mathrm{2D}$.
For the imbalanced pump,  we found additionally (V) a   superfluid striped phase with light scattering coupled to the cavity $P$ quadrature SR2+SF$_\mathrm{1D}$ and (VI) an insulating striped phase with light scattering coupled to the cavity $P$ quadrature SR2+MI$_\mathrm{1D}$. 
 In both cases, increasing $\Delta_c$ (VII), an  unstable phase appears, where  there is no stationary solution to $\alpha$. 

Our results from simulations are consistent with previously reported experimental observations in the absence of atomic interactions, supporting only superfluid states~\cite{PhysRevResearch.3.L012024, Dreon2022,PhysRevLett.123.233601}. Beyond the non-interacting case, in the strongly interacting regime,  the presence of atomic collisions (Hubbard $U$) triggers  the formation of Mott insulator regions  in the phase diagram, originated in the suppression of the atomic on-site density fluctuations  $\overline{\Delta(\hat n)^2}$ and the condensate fraction~\cite{Burnett,book:92350169} $f_c$ due to strong quantum correlations.
This occurs in spite of  the absence of potential barriers between the sites,  as the repulsive potential is strong enough to localize the  Wannier functions in both directions. This behavior is specific to the blue detuned case
due to the contribution of the  $Q$ quadrature interference term. 

Characterizing the atomic behavior, the  static structure factor can be directly measured and is defined as the fluctuation  $   S(\textbf{q})= (\braket{ \hat\rho_\textbf{q} \hat\rho_{-\textbf{q}}}-| \braket{ \hat\rho_\textbf{q} }|^2)/N$, where   the $\textbf{q}$ component of the density operator is $  \hat \rho_\textbf{q}$$=\int d\textbf{r} e^{-i\textbf{q}\cdot\textbf{r}} \hat n(\textbf{r})$\cite{PitaevskiStringari}.
 In  Fig. \ref{fig:Fig1} e-g, i-k, the maximum of the static structure factor in the first Brillouin zone is plotted along the lines of the phase diagrams of Fig. \ref{fig:Fig1} d, h, along with the radial component of the minimum $\textbf{q}_{r,\mathrm{max}}$, the condensate fraction, the mean of the particle number fluctuations per site, the structural order parameters $|\Phi_x|$, $|\Phi^{(1)}_{xy}|$ and $|\Phi^{(2)}_{xy}|$, and the number of cavity photons $N_\mathrm{ph}=|\alpha|^2$. It can be seen that, when there is no light scattered into the cavity (phase I in Fig.\ref{fig:Fig1} e, f and i), $\textbf{q}_\mathrm{max}$ is a non zero vector located  inside the first Brillouin zone. As the system enters to a superradiant phase ($N_\mathrm{ph}\gg 0$), the maximum is displaced to a corner of the first Brillouin zone, where $q_{r,\mathrm{max}}$ adopts its maximum value. In the MI (phases IV and VI in Fig.\ref{fig:Fig1} f, g, j and k), the condensate fraction, the structure factor and  the atom number fluctuations  are strongly suppressed. Similar to classical structural phase transitions at finite $T$, we find that quantum phase transitions of weak first order, first order and second order emerge while the systems changes dimensionality and the phase of quantum matter changes.

\begin{figure}
    \centering
    \includegraphics[width=0.49\textwidth]{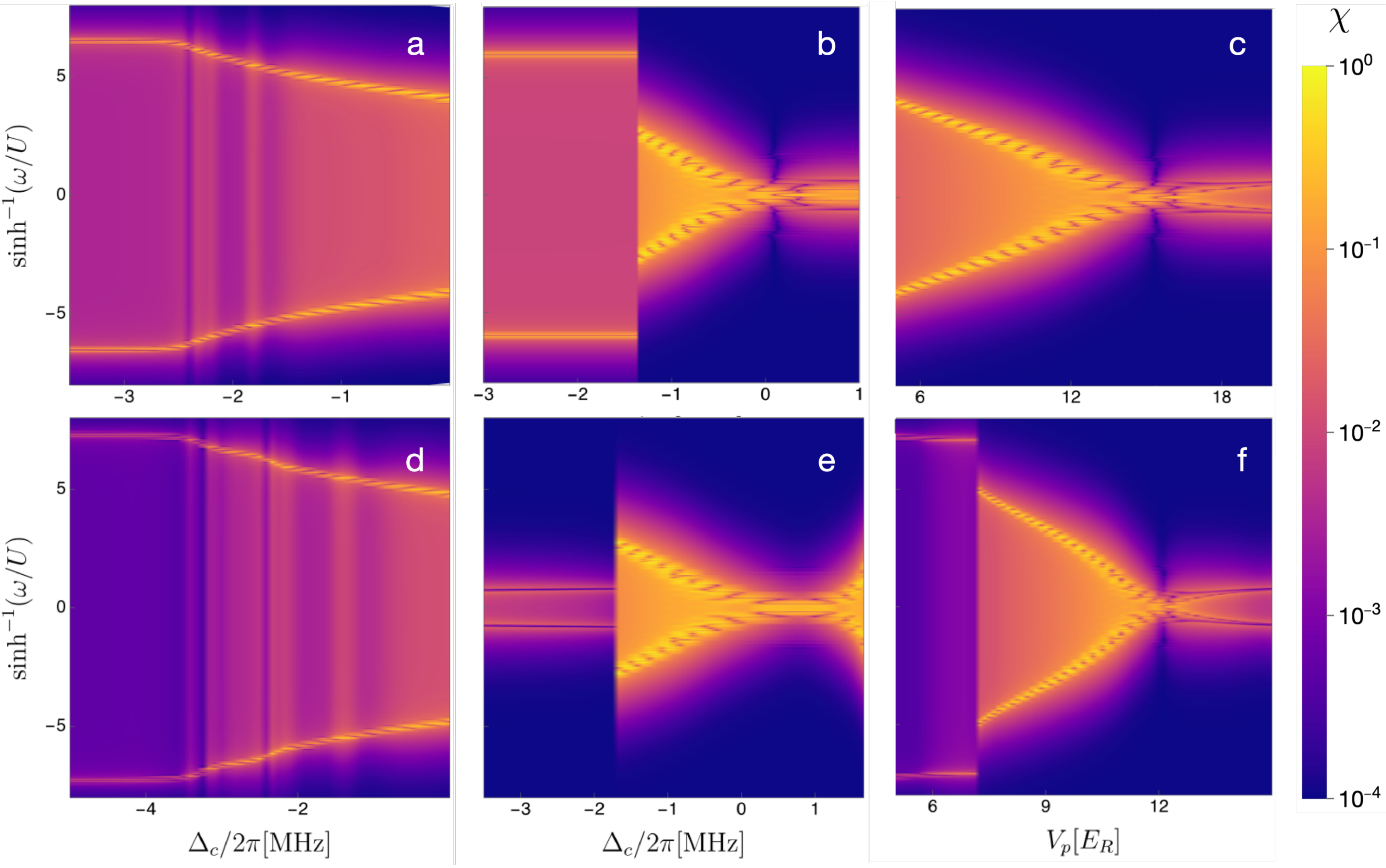}
    \caption{
     \textbf{Mean dynamic polarizability ($\chi$) for different trajectories across quantum phase transitions}.
     Panels (\textbf{a},\textbf{b},\textbf{c},\textbf{d},\textbf{e},\textbf{f}) correspond to the trajectories  (\textbf{e},\textbf{f},\textbf{g},\textbf{i},\textbf{j},\textbf{k}) in  Fig.\ref{fig:Fig1} with the same parameters. Weak mode softening can be observed in \textbf{a} and \textbf{d} after first order transitions from the non radiant SF state to the superradiant superfuid in 1D. The lattice geometry (1D or 2D)  triggered by the light inside the cavity controls how the quantum phase transition points are approached either as second order when the structure is preserved or first order when there is a change in geometry. Panels \textbf{b},\textbf{c},\textbf{e},\textbf{f}:  when the underlying geometry of the lattice (1D or 2D)  is preserved towards the MI superradiant state with either SR1 or SR2 emision, the system presents mode softening, characteristic of a second order phase transition.
    }
    \label{fig:Fig2}
\end{figure}

The nature of the phase transitions observed can be further analyzed using  the dynamical structure factor\cite{PitaevskiStringari} $S(\mathbf{q},\omega)$, which is measurable~\cite{Landig2015}.
Access to the dynamic structure factor  allows  to characterize the relevant quasi particle mode behaviour via the dynamic polarizability while the system crosses a structural phase transition.  The dynamic polarizability is related to $S(\mathbf{q},\omega)$  via~\cite{PitaevskiStringari,PhysRevA.79.043607,Nozieres},
\begin{equation}
\label{pol}
\chi_\mathbf{q}(\omega)=\int_0^\infty \frac{2\omega' S(\mathbf{q},\omega')}{\omega'^2-\omega^2}\mathrm{d}\omega'.
\end{equation}
We estimate $S(\mathbf{q},\omega)$ from our self-consistent light-matter DMRG simulations using Krylov methods~\cite{RevModPhys.77.259}, see End Matter. We observe the collective mode behaviour with  the mean dynamic polarizability $\chi(\omega)=\int_{\mathrm{BZ}} \chi_{\mathbf{q}}(\omega)\mathrm{d}^2q$, integrated over the Brillouin zone ($\mathrm{BZ}$)  in Fig.\ref{fig:Fig2}. Twin peaks of high frequency modes are the typical signal in the SF state. In the MI the twin peaks are strongly suppressed in amplitude and the frequency separation is strongly reduced. Two aspects explain this behavior, one is the amplitude of the static structure factor that is minimal in the insulating state and the other is the fact that as the gap opens in the MI, the ground state degeneracy is suppressed. Effectively the number of states contributing  the signal in $\chi$ is maximized in the SF state. Moreover, the combination with the dimensionality of the lattice naturally suppresses the number of nearly energy equivalent states for the anisotropic states. Thus, the number of excited states contributing to the signal of $\chi$ in the 2D geometry is larger than in 1D.
The transitions from the normal superfluid with anisotropy to other states are always first order, signaled by the change in the slope in the trajectory of the collective modes in $\chi$. The SR1 states present weak mode softening changing dimensionality (first order), there is a weak shift of the peaks in $\chi$ towards the zero fequency. However, when the geometry is 2D in the superradiant phase in the transition  between the SF and the MI estate,  we find bonafide mode softening typical of a second order quantum phase transition.  Here the collective mode amplitudes approach the zero frequency when reaching the transition point from either side of the transition. Interestingly, transitions from the SR2 insulator states (SR2-MI$_\mathrm{1D}$) can be of any kind depending if there is change in dimensionality. If anisotropy is preserved the transition is second order, while if the lattice geometry becomes 2D then the transition is first order. In general, it can be seen that if the lattice does not change structure the quantum phase transitions towards the SR insulating states are second order. Thus, the light build up in the cavity selects the character of the structure in the lattice and the nature of the quantum phase transitions between different phases of quantum matter.

The quotient $ t/U$, between the effective tunneling amplitude $t$ and the effective atomic interaction strength  $U$, can be controlled by 
tuning the pump optical lattice depth $V_p$ and the dynamical optical lattice contribution $V_c\propto\langle\hat a^\dagger\hat a\rangle$ (see End Matter) controlling the localization of atomic orbitals. Alternatively,  
it is possible to modify the value of the coupling constant $g_\mathrm{2D}$ through changing the oscillator length of the confinement in the perpendicular direction from the plane of the optical lattice~\cite{PhysRevLett.84.2551} or by tuning $a_s$ using Feshbach resonances~\cite{RevModPhys.82.1225}.    
For $^{87}$Rb gases, as used in \cite{PhysRevLett.123.233601, PhysRevResearch.3.L012024,Landig2015}, the widths of the known Feshbach resonances are small ($<0.22$ G)\cite{RevModPhys.82.1225}, making difficult  to tune  $a_s$, since it requires a great degree of precision in the control of the external magnetic field. A new generation of experiments using $^{39}$K  will open up the possibility of overcoming this issue, since the Feshbach resonances widths for these atomic species are broader\cite{PhysRevA.81.032702} and experimentally viable to manipulate.

In conclusion, we have developed a self-consistent theoretical framework to the describe light and quantum matter in the strongly correlated regime. We have studied the properties of quantum matter in a realistic setup motivated by recent experimental realizations of ultracold atoms pumped by a blue detuned optical lattice, coupled to the two orthogonal quadratures in an optical cavity. We have found several new insulating phases of quantum matter that are present in the limit beyond weak atomic interactions. We analyzed via simulations the emergence and properties of the transitions between the superradiant and normal phases in the system. We could corroborate the properties of the transitions found using the structure factor and polarizability which are experimentally accessible. Interestingly, we were able to recover key features of the transitions such as mode softening from our light-matter DMRG simulations.  Our findings can be explored with current ultracold experiments and provide additional information to understand the physics of structural phase transitions in the quantum limit. The tools and techniques developed in this work can be used to model and analyze the properties of light-matter quantum systems in general and other state of the art cavity experiments such as two level systems\cite{Rey}, multi-level atoms\cite{Monika} and ultracold fermions\cite{JPhillippe}.

\begin{acknowledgements}
We thank  R. Rosa-Medina, P. Christodoulou, T. Donner, F. Mivehvar and H. Ritsch for helpful discussions. This work is partially  supported by the grants DGAPA-UNAM-PAPIIT: IN118823 and CONAHCYT-CB:A1-S-30934.
\end{acknowledgements}

\bibliography{biblio}
\clearpage
\newpage

\section{End matter}
The many body Hamiltonian of the system rotating at frequency $\omega_p$ is \cite{Maschler2008,NewJPhys.17.123023}
\begin{equation}\label{eq:HamIni}
    \hat H= \hat H_{\mathrm{ph}} + \hat H_{\mathrm{kin}}+ \hat H_{\mathrm{pump}} +\hat H_{\mathrm{cav}} + \hat H_{\mathrm{pump-cav}} +  \hat H_{\mathrm{int}},
\end{equation}
where $\hat H_{\mathrm{ph}}=-\hbar\Delta_c\hat a^\dagger\hat a$ is the energy of the cavity photons, $\Delta_c = \omega_p-\omega_c$ is the cavity detuning, $\hbar$ the reduced Planck constant, $\hat a^{(\dagger)}$ is the cavity mode photonic annihilation (creation) operator, $\hat H_{\mathrm{kin}}$ is the kinetic energy of atoms, $\hat H_{\mathrm{pump}} =  V_p\hat W$,  $\hat H_{\mathrm{cav}} =  U_0\hat a^\dagger \hat a\hat B$ and  $   \hat H_{\mathrm{pump-cav}} =  f_1 \sqrt{U_0 V_p} (\hat a +\hat a^\dagger)\hat \Theta_1-i f_2\sqrt{U_0 V_p}(\hat a -\hat a^\dagger)\hat \Theta_2$ are the pump,  cavity  and  pump-cavity interference potential terms, respectively. 
The operators 
$\hat W =\int d^2 r \hat \Psi^\dagger(\textbf{r}) \cos^2(\textbf{k}_p\cdot\textbf{r})\hat \Psi(\textbf{r}) $, $\hat B=\int d^2 r \hat \Psi^\dagger(\textbf{r}) \cos^2(\textbf{k}_c\cdot\textbf{r})\hat\Psi(\textbf{r})$,  with $\hat\Theta_1$ and $\hat\Theta_2$ determine the spatial profile of the optical lattice. The imbalance parameter $\gamma$ determines the constants
$f_1 = (\gamma + \gamma^{-1})/2$ and $f_2 = (\gamma - \gamma^{-1})/2$, which modulate the coupling  to the $\hat Q =(\hat a + \hat a^\dagger)/\sqrt{2}$ and  $\hat P =i(\hat a^\dagger - \hat a)/\sqrt{2}$ quadratures of the cavity field.
 $V_p$ denotes the pump lattice depth, $U_0>0$ the dispersive shift per atom and $V_c=U_0\braket{\hat a^\dagger \hat a}$ the emerging cavity lattice depth.
$\hat H_\mathrm{int}=g_\mathrm{2D}\int d^2 r \hat \Psi^\dagger(\textbf{r}) \Psi^\dagger(\textbf{r})\Psi(\textbf{r})\hat\Psi(\textbf{r})$ is the two body  interaction term, with $g_\mathrm{2D}$ the two dimensional coupling constant. 
Assuming harmonic confinement along the direction $z$ orthogonal to the system and $l_z>> |a_s|$, where $l_z$ is the oscillator length along the $z$ direction, the interaction strength is $g_\mathrm{2D}=2\sqrt{2\pi}\hbar^2 a_s /(m l_z)$ \cite{PhysRevLett.84.2551}, where  $a_s$ is the  $s$-wave scattering length.

Due to the different time scales of the atomic and the cavity field dynamics, one can adiabatically eliminate the latter\cite{Maschler2008}.
The photonic operator $\hat a$ is replaced by its  steady state value $\alpha=\braket{\hat a}$, which is computed from the static Heisenberg equation $0=d\braket{\hat{a}}/dt = (i/\hbar)\braket{[\hat{H},\hat{a}]}- \kappa\braket{\hat{a}}$,   where dissipation from the cavity is introduced phenomenologically through the cavity decay rate $\kappa$, obtaining
\begin{equation}\label{eq:alpha}
    \alpha =\frac{\sqrt{U_0 V_p}}{\hbar\Delta_c -  U_0\braket{\hat B}+i\hbar \kappa}(f_1\braket{\hat \Theta_1} + i f_2\braket{\hat \Theta_2}),
\end{equation}
which is solved self consistently as function of $V_p$ and $\Delta_c$.
With the aim of exploring the role of strong atomic interactions,   the atomic field operators are expanded as $\hat \Psi^{(\dagger)}(\textbf{r})=\sum_{i} \hat b_{i}^{(\dagger)} w^{(*)}(\textbf{r}-\textbf{r}_i)$, where
$\hat b_{i}^{(\dagger)}$  annihilates (creates) a particle  in the Wannier state $w(\textbf{r}-\textbf{r}_i)$ of the lowest band localized  at the site $\textbf{r}_i$. 
In the $\Delta_a>0$ case, the lattice potential emerging form the coupling to the $Q$ quadrature is not formed by  potential wells. Furthermore, the lattice  potential  is not isotropic. This makes non trivial  to approximate  the Wannier functions analytically.
Therefore, we use the band projected position operator method \cite{PhysRevLett.111.185307,Kivelson,PhysRevB.103.075125} to compute them dynamically as a function of the light in the cavity mode.  The  effective Bose Hubbard Hamiltonian with light dependent coefficients via the Wannier functions is, 
\begin{equation}\label{eq:HamLight}
    \begin{split}
    \hat H(\alpha) = & \epsilon (\alpha)\sum_i   \hat n_i-\sum_{i\neq j} t_{ij}(\alpha) \hat b^\dagger_i \hat b_j 
     \\
     & +\frac{U(\alpha)}{2}\sum_i\hat n_i(\hat n_i-1)-\hbar\Delta_c|\alpha|^2,
    \end{split}
\end{equation}
where 
$\hat b_{i}^{(\dagger)}$  annihilates (creates) a particle in the Wannier state of site $i$ of the lowest band and $\hat{n}_i =\hat b_{i}^{\dagger}\hat b_{i} $. The tight binding coefficients are dynamically linked to the light field $\alpha = \braket{\hat a}$ via the Wannier functions. 
 The tunneling coefficients $t_{ij}(\alpha) = -\int d^2 r w_\alpha^*(\textbf{r}-\textbf{r}_i)\left[ -\frac{\hbar^2}{2m}\nabla^2 + V_\mathrm{OL}(\alpha,\textbf{r})   \right] w_\alpha(\textbf{r}-\textbf{r}_j)$ are given by the overlap of Wannier functions at sites $i$ and  $j$ and  the kinetic and  lattice potential terms of the Hamiltonian. The effective mean field optical lattice potential is $ V_\mathrm{OL}(\alpha,\textbf{r})=V_p \cos^2(\textbf{k}_p\cdot\textbf{x})  + U_0|\alpha|^2 \cos^2(\textbf{k}_c\cdot\textbf{x}) + 2\sqrt{U_0 V_p}[f_1\mathrm{Re}(\alpha) \cos(\textbf{k}_p\cdot\textbf{x})\cos(\textbf{k}_c\cdot\textbf{x})+f_2\mathrm{Im}(\alpha) \sin(\textbf{k}_p\cdot\textbf{x})\cos(\textbf{k}_c\cdot\textbf{x}) ]$.

The on-site energy $\epsilon(\alpha)=-t_{i,i}(\alpha)$ and 
the on site interaction coefficient  $ U(\alpha)=g_\mathrm{2D}\int d^2 r  |w_\alpha(\textbf{r}-\textbf{r}_i)|^4$
are site independent.

The on site  and the cavity field energy are present on the model, as the total energy is needed to determine if light scattering in the cavity mode is energetically favorable.

For our simulations, we use  relevant experimental parameters\cite{PhysRevLett.123.233601}: 
 $N=2.7\times10^{5}$ is the number of atoms, $U_0 = 0.012 \ E_{R}$ and  $\kappa=2\pi\times147$ kHz.  The recoil energy   $E_R = (\hbar k_p)^2/(2m) = 2\pi\hbar\times3.77$ kHz is used as our energy scale.  The cavity decay rate $\kappa$ is safely neglected, as this leads to a minor shift $\delta\theta_\alpha =\arctan[-\kappa/(\Delta_\mathrm{eff})]$ on the complex phase $\theta_\alpha\equiv\mathrm{arg}(\alpha)$ of the cavity field, where  $\Delta_\mathrm{eff}=\Delta_c -U_0\braket{\hat B}/\hbar$ is the effective cavity detuning.

The Bose Hubbard model at zero temperature describes a quantum phase transition between a superfluid  and a Mott Insulator \cite{book:92350169}, driven by the competition between tunneling processes and on site atomic interactions, characterized by the quotient $\bar t/U$, where $\bar t$ is the effective tunneling amplitude.
From the ground state and excited states of the Bose-Hubbard model at commensurate density per site $\bar n=1$ obtained from  our light-matter DMRG calculations we obtain the results in Figs. \ref{fig:Fig1} and \ref{fig:Fig2} .

\subsection*{Self Consistent Algorithm: light-matter DMRG}\label{sec:SelfConsistentAlgorithm}

The phase diagrams are obtained solving \eqref{eq:alpha} self consistently. Giving a ansatz $\alpha\neq 0$, we obtain the ground state of $H(\alpha)$ and obtain a new value for $\alpha$\ from \eqref{eq:alpha}. This cycle is repeated until convergence is achieved with a tolerance less than single precision. There can be multiple solutions for $\alpha$ for fixed $V_p$ and $\Delta_c$, so it is necessary to keep only the lowest energy  state.
When  interactions are neglected in the model, the simplest approach to finding the ground state is to expand the field operators in the plane wave basis $\hat\Psi(\textbf{r}) = A^{-1/2}\sum_\textbf{k}e^{i\textbf{k}\cdot\textbf{r}}\hat b_\textbf{k}$. For the interacting case, to obtain the ground state  of \eqref{eq:HamLight} given $V_p$ and $\alpha$, one needs to compute the tight binding coefficients using  the Wannier functions  of the full pump-cavity optical lattice at each step. To find the effective ground state of (\ref{eq:HamLight}) for a fixed parameter set in the self-consistent loop, we perform simulations using DMRG~\cite{DMRG1,RevModPhys.77.259,DMRG3} with the aid of the ITensor library\cite{itensor1, itensor2}. We build a 2D square lattice with a torus geometry with 7  by 14 (98) sites to resemble full periodic boundary conditions based on~\cite{DMRG3} and minimize finite size effects; we verified lower number of sites recover qualitative similar results. In addition,  we compute up to 80 excited states using Krylov methods and determine the  continued fraction representation of $S(\omega,\mathbf{q})$ following~\cite{RevModPhys.77.259} page 282; we verified similar qualitative results are recovered for $20\%$ less excited states. The maximum number of atoms considered per site is $n_{\textrm{max}}= 4$.


\clearpage
\newpage

\widetext
\begin{center}
\textbf{\large Supplemental material for: Structural Dynamics and Strong Correlations in Dynamical  Quantum Optical Lattices}
\end{center}
\setcounter{equation}{0}
\setcounter{figure}{0}
\setcounter{table}{0}
\setcounter{page}{1}
\makeatletter
\renewcommand{\theequation}{S\arabic{equation}}
\renewcommand{\thefigure}{S\arabic{figure}}

\section*{Effective Hamiltonian }

When the lowest energetic band of the optical lattice band spectra is a composite band, there are several Wannier functions localized at different sites on each unit cell.
A rigorous expansion in the Wannier basis would include all the intersite couplings and on site terms, which increases the complexity of the analysis.  
To simplify the model and limit the number of terms appearing in the Hamiltonian,  we introduce an effective Wannier function per unit cell, as the superposition that minimizes the total energy in the non interacting case of the Wannier functions belonging to each unit cell. 
The effective  Hamiltonian is
\begin{equation}\label{eq:HamBH}
    \hat H(\alpha) =  \epsilon (\alpha)\sum_i   \hat n_i-\sum_{i\neq j  }t_{ij}(\alpha) \hat b^\dagger_i \hat b_j 
      +\frac{U(\alpha)}{2}\sum_i\hat n_i(\hat n_i-1)-\hbar\Delta_c|\alpha|^2,
\end{equation}
where the on site amplitude $\epsilon(\alpha)$ and the interaction coefficient $U(\alpha)$ become site independent.
We do not restrict the sum in the second term of \eqref{eq:HamBH} to nearest neighbor couplings, since due to the absence of energetic barriers for  the blue detuned optical lattice, the Wannier functions become broader compared to as if they were localized inside a potential well.
The necessity of including terms beyond  nearest neighbors can be  verified comparing the non interacting phase diagrams in the momenta and site representations in the decoupling approximation\cite{book:92350169}. To simplify \eqref{eq:HamBH}, we replace the tunneling amplitudes by  an effective tunneling amplitude $t$, which we define as the mean of all the tunneling amplitudes $\bar{t}(\alpha) \approx(N_s-1)^{-1} \sum_{j\neq i} t_{ij}(\alpha)$, rendering effectively site independent and  only between nearest neighbors.

\section*{Structure factor and Dynamic Polarizability}
The static structure factor $S(\textbf{q})$ is defined as the fluctuactions of the 
  Fourier $\textbf{q}$-component  $\hat\rho_\textbf{q}$ of the density operator $\hat \rho(\textbf{r}) = \hat \Psi^\dagger(\textbf{r})  \hat \Psi(\textbf{r})$,

    $S(\textbf{q})=\frac{1}{N} (\braket{ \hat\rho_\textbf{q} \hat\rho_{-\textbf{q}}} -| \braket{ \hat\rho_\textbf{q} }|^2)$,

where $\hat \rho_\textbf{q}=\int d\textbf{r} e^{-i\textbf{q}\cdot\textbf{r}} \hat \rho(\textbf{r})$,  $\hat \Psi(\textbf{r})$ is the atomic field operator and $N$ is the number of atoms.
Expanding the field operator $\hat \Psi(\textbf{r})$ in the Wannier basis and introducing the Fourier transform $    \mathcal{W}_{i,j}(\textbf{q})= \int d\textbf{r}  e^{-i\textbf{q}\cdot\textbf{r}} w^*(\textbf{r})w(\textbf{r} - \textbf{a}_{ij})$ of the product of Wannier functions, where $\textbf{a}_{ij}=\textbf{r}_j-\textbf{r}_i$, one arrives to 
\begin{equation}\label{eq:StaticStructureFactorComplete}
    S(\textbf{q})= \frac{1}{N} \sum_{i,j,k,l}  e^{i\textbf{q}\cdot(\textbf{r}_k-\textbf{r}_i)} \mathcal{W}_{i,j}(\textbf{q}) \mathcal{W}_{k,l}(-\textbf{q}) C(\hat b^\dagger_i \hat b_j ,\hat b^\dagger_k  \hat b_l ), 
\end{equation}
where the quantum covariance of two operators is defined as $C(\hat A, \hat B) = \braket{\hat{A
}\hat{B}}-  \braket{\hat{A
}}\braket{\hat{B}}$ with the expectation value in the ground state. \eqref{eq:StaticStructureFactorComplete} is used to compute the maximum of the structure factor, shown in Fig. \ref{fig:Fig1} of the main text. The dynamic structure factor can be computed by using the methods in~\cite{RevModPhys.77.259}. Using the dynamic structure factor in continued fraction representation, we compute the  dynamic polarizability via the formula in the main text \eqref{pol} \cite{PitaevskiStringari,PhysRevA.79.043607,Nozieres}.

\begin{figure}
    \centering
   \includegraphics[scale=0.8]{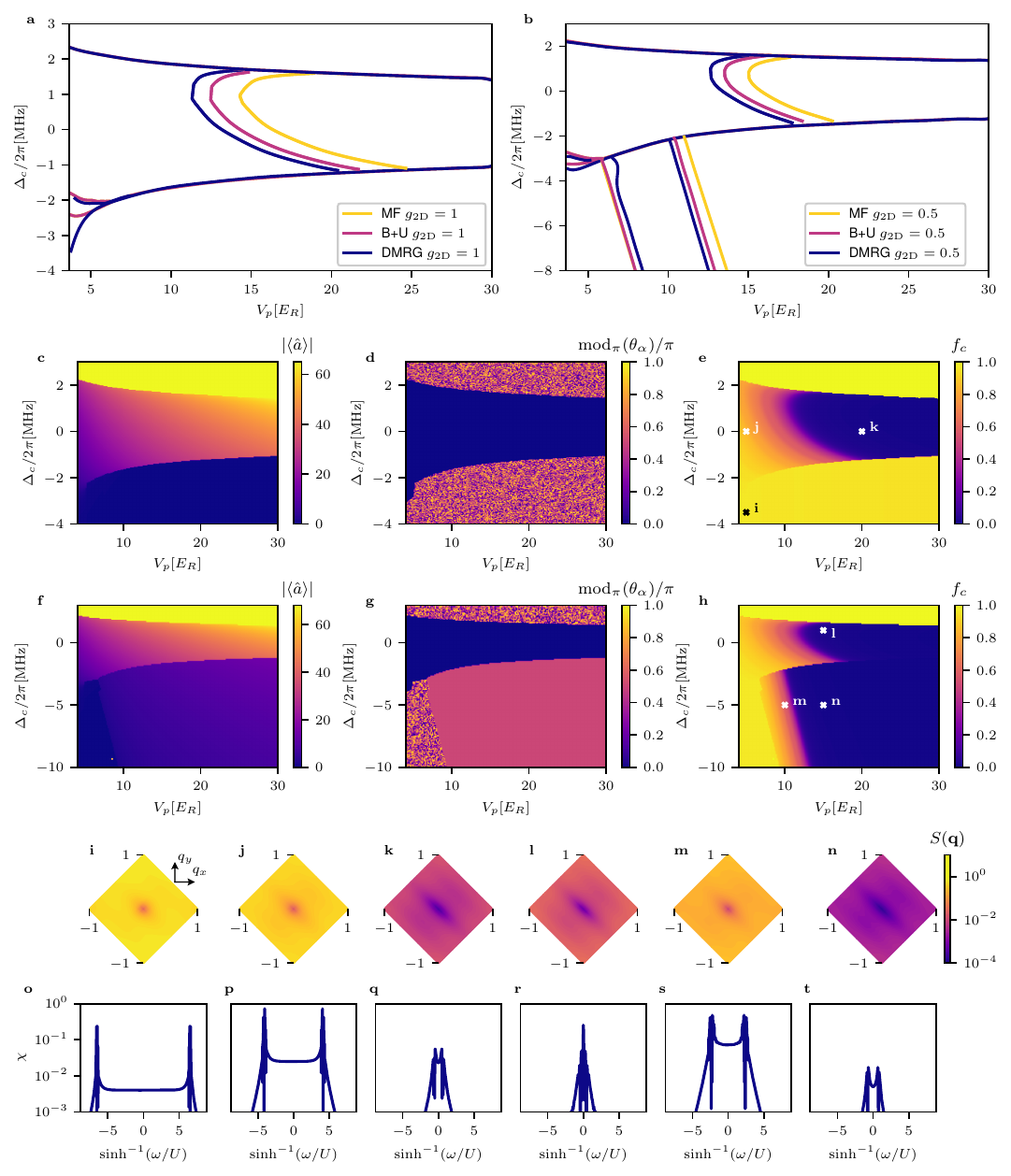}
    \caption{\textbf{Phase Diagrams for different interaction strengths.}
    \textbf{a}, \textbf{b}, Phase boundaries  obtained by the decoupling (MF) and  B+U  approximations and by the DMRG. 
    We use  (\textbf{a}) $\gamma=1$, $g_{2D}=1.0$  and (\textbf{b}) $\gamma=1.37$, $g_{2D}=0.5$, for the balanced and imbalanced case, respectively.  \textbf{c}, \textbf{f}, Light amplitude $|\alpha|=|\braket{\hat a}|$. \textbf{d}, 
    \textbf{g}, Complex light phase modulus $\pi$, $\mathrm{mod}_{\pi}(\theta_\alpha)/\pi$. \textbf{e}, \textbf{h}, Condensate  fraction $f_c$. 
     We use $\gamma=1$, $g_{2D}=1.0$ (\textbf{c}-\textbf{e}) and $\gamma=1.37$, $g_{2D}=0.5$ (\textbf{f}-\textbf{h}). \textbf{i}-\textbf{n},
     Static structure factor in the first Brillouin zones  for the different points marked in (\textbf{e}) and (\textbf{h}). The logarithmic colorbar scale is at the right of \textbf{n}.
      \textbf{o}-\textbf{t}, Mean dynamic polarizability, for the points of the phase diagram corresponding to the plots (\textbf{i}-\textbf{n}).
     }
    \label{fig:supp1}
\end{figure}

\section*{Phase diagrams}
In Fig. \ref{fig:supp1}a, b, we show the phase boundaries for  the phase diagrams obtained in the main text (Fig. \ref{fig:Fig1 }d, h). The boundaries where obtained from the decoupling\cite{book:92350169} and B+U \cite{PhysRevB.91.224510} approximations, as well as, from light-matter DMRG calculations. 
In the DMRG calculations we define the SF-MI phase boundary as the point where the curvature of the condensate fraction is maximized, as function of $t/U$ consistent with finite scaling analysis.
In Fig. \ref{fig:supp1}c-h, we plot the light amplitude (\ref{fig:supp1}c, f), the complex light phase (\ref{fig:supp1}d, g) and the condensate fraction (\ref{fig:supp1}e, h) for the balanced \ref{fig:supp1}(c-e) and imbalanced (\ref{fig:supp1}f-h) cases, using the DMRG scheme.
These three quantities function as order parameters for the different phases of the system. The superradiant phase boundary is defined as the point where the second derivative of $|\alpha|^2$ with respect to $\Delta_c$ is maximized or not defined, by a discontinuity in $|\alpha|^2$. The complex phase of the light field allows us to distinguish between the two superradiant phases, where $\mathrm{mod}_{\pi}(\theta_\alpha)/\pi = 0$ for SR1 and  $\mathrm{mod}_{\pi}(\theta_\alpha)/\pi = 1/2$ for SR2.  
In Fig. \ref{fig:supp1}i-n we plot the static structure factor in the first Brillouin zone, for the marked points in 
Fig. \ref{fig:supp1}e, h. We found that the zero component of the static structure factor vanishes in all cases. The static structure factor is maximal deeper into the superfluid phase, while it decreases significantly entering the insulator phase, since the fluctuations of the density in the insulator phase are strongly suppressed. 


\end{document}